\documentclass[a4paper,11pt]{article}
\pdfoutput=1 
\usepackage{jheppub} 

\usepackage[T1]{fontenc} 

\newcommand{\pd}{\partial}

\newcommand{\nn}{\nonumber}

\newcommand{\vp}{\varphi}

\newcommand{\cG}{\mathcal{G}}
\newcommand{\cF}{\mathcal{F}}

\def\K{K{\"a}hler}
\def\be{\begin{equation}}
\def\ee{\end{equation}}
\newcommand{\ba}{\begin{eqnarray}}
\newcommand{\ea}{\end{eqnarray}}

\newcommand{\rf}[1]{(\ref{#1})}

 \title{\rm {\bf \huge  \boldmath $\overline {D3}$ Induced Geometric Inflation}}

\author[a]{Renata Kallosh,}
\author[a]{Andrei Linde,}
\author[b]{Diederik Roest,}
\author[a]{and Yusuke Yamada}
\affiliation[a]{Stanford Institute for Theoretical Physics and Department of Physics, Stanford University, Stanford, CA 94305, USA}
\affiliation[b]{Van Swinderen Institute for Particle Physics and Gravity, 
University of Groningen, Nijenborgh 4, 9747 AG Groningen, The Netherlands}
\emailAdd{kallosh@stanford.edu}
\emailAdd{alinde@stanford.edu}
\emailAdd{d.roest@rug.nl}
\emailAdd{yusukeyy@stanford.edu}

\notoc

\abstract{Effective supergravity inflationary models induced by anti-D3 brane interaction with the moduli fields in the bulk geometry have a geometric description. The \K\, function  carries the complete geometric information on the theory. The non-vanishing bisectional curvature plays an important role in the construction. The new geometric formalism, with the nilpotent superfield representing the anti-D3 brane, allows a powerful generalization of the existing inflationary models based on supergravity. They can easily incorporate arbitrary values of the Hubble parameter, cosmological constant and gravitino mass. We illustrate it by providing generalized versions of polynomial chaotic inflation, T- and E-models of $\alpha$-attractor type,  disk merger. We also describe a multi-stage cosmological attractor regime, which we call cascade inflation.}

\begin{document}

\maketitle

\newpage

 \tableofcontents{}

\section{Introduction}

\parskip 3pt

In the KKLT scenario for moduli stabilization~\cite{Kachru:2003aw}, spontaneous supersymmetry breaking can be induced by an anti-D3-brane in the Calabi-Yau bulk geometry. Its worldvolume theory includes a Volkov-Akulov fermion goldstino~\cite{VA}, which can be equivalently described in terms of a nilpotent scalar superfield $S$ with $S^2(x, \theta)=0$~\cite{nilpotent,Nonlinear}. This nilpotent superfield allows for a manifestly supersymmetric description of the uplift to a de Sitter minimum~\cite{Kallosh:2014wsa,Kallosh:2015nia}. In a parallel development, it has been realized that nilpotent superfields have great potential as a model building tool in effective supergravity theories of inflation~\cite{NilpotentInflation}. The nilpotent multiplet helps to achieve stability of inflationary trajectory, making the non-inflaton fields of the theory heavy. 

One would like to connect these two developments: can the $\overline{D3}$-brane interactions with the Calabi-Yau moduli give rise to effective supergravity descriptions of inflation? It is not known how the $\overline{D3}$-brane interacts with the CY moduli fields $(T^i, \overline{ T}^i)$. However, one may ask a question: what kind of interaction between $S$ and $(T^i, \overline{ T}^i)$ would lead to phenomenological supergravity models of inflation, including the exit stage, that are compatible with the data? 

Here we will construct what we call  $\overline {D3}$ induced geometric inflation models.
In these models, once one decides about the potential $\mathbf{ V} (T^i, \overline{ T}^i)$, it is easy to find the corresponding $S$-field geometry  $\cG_{S\overline S} (T^i, \overline{ T}^i)$  in the supergravity \K\,  function $\cG$, and one is guaranteed to reproduce the desired potential during inflation. However, one has still to check the stability of each model and show the absence of tachyons. The bisectional curvature of these geometric models  will play a role in the stability analysis.  

We will develop a general class of $\overline {D3}$ induced geometric inflation with multiple moduli in CY bulk interacting with $\overline {D3}$ nilpotent multiplet $S$. It is  important that the  $\overline {D3}$ induced geometric inflation models have a non-vanishing gravitino mass -- $W$ does not vanish during and at the exit from inflation. In this case, one can use the advantage of a geometric \K\, function formalism where
\be
\cG \equiv K + \log W +\log \overline W\, , 
\qquad 
    \mathbf{ V} = e^ {\cal   G}  (\cG^{\alpha \overline \beta }  {\cal   G} _\alpha  {\cal   G} _{\overline \beta}  - 3 )  
\ee
and study various interesting application of the new models. Here the index $\alpha$ includes the directions   $ S$ and $ T^i $.

The role of the \K\, function $\cG$ was recognized starting with~\cite{Cremmer:1978iv} when  supergravity models interacting with matter were first constructed. It was shown there that the action is fully determined by the \K\, function. 
However, in some cosmological models, for example in D-term inflation~\cite{Binetruy:1996xj}, or in models in~\cite{Kallosh:2010xz}, during the evolution the superpotential might vanish.  For these models it was more useful to employ the \K\, potential  and the superpotential $W$ since the \K\, function $\cG$ has a singularity at  $W=0$.  Meanwhile,  the  analysis of metastable de Sitter vacua with spontaneously broken supersymmetry was based mostly on the analysis using the \K\, function $\cG$, see for example,~\cite{Covi:2008cn}. Comparative to this analysis, the new ingredient here is the fact that the $S$ superfield is nilpotent and that we will use it  for developing inflationary models with the exit to de Sitter minima. Our Hermitian \K\, function  will be of the form
\be
\cG (T^i, \overline{ T}^i; S, \overline S) = \cG_0(T^i, \overline{ T}^i) + S +\overline S + \cG_{S\overline S} (T^i, \overline{ T}^i)S\overline S \,,
\label{cG0}\ee
which we will show will describe the general case of supergravity models with one nilpotent multiplet and  non-vanishing superpotentials.

We will show below that, in general, from the knowledge of the potential $\mathbf{ V} (T^i,\overline{ T}^i)$ and the $T$-dependent \K\, function $\cG_0 (T^i, \overline{ T}^i)$ it is possible to recover the $S$-field geometry
\be
\cG_{S\overline S} (T^i, \overline{ T}^i) dS d\overline S.
\label{metric}\ee
Whereas the complete formula will be given below in eq.~\rf{general}, here we would like to point out that under certain conditions 
 the relation between the $S$-field geometry and the potential simplifies significantly.  If the gravitino mass is constant throughout inflation at $S=0$, and supersymmetry is unbroken in the $T^i$ directions, i.e. during inflation we have
 \be
e^{ {\cal   G} (T^i, \overline{ T}^i)}=|m_{3/2}|^2={\rm const} \,, \qquad 
 \cG_{T^i} (T^i, \overline{ T}^i)=0\, , \label{K}\ee
one finds  the following simple relation between the inflationary potential and the geometry:
 \begin{align}
\cG_{S \overline S} (T^i, \overline{ T}^i)  =   {  |m_{3/2}|^2\over \mathbf{ V} (T^i,\overline{ T}^i)+3 |m_{3/2}|^2 }    .
 \label{new}  \end{align} 
Here  $\mathbf{ V} (T^i, \overline{ T}^i)$ is the scalar potential of supergravity at $S=0$ defined in the standard way either from the \K\, function $\cG$ or from the superpotential $W$  and \K\, potential $K$. 
In  examples of $\overline {D3}$ induced geometric inflation models which we will specify below, the conditions \rf{K} will be satisfied during and after inflation.

Examples of models with non-trivial Hermitian function $\cG_{S\overline S} (T, \overline T)$ include warped Calabi-Yau throats~\cite{Kallosh:2015nia}, in which the \K\, potential takes the form
\be
-\ln (T+\overline T - S\overline S)= -\ln (T+\overline T ) + { S\overline S\over T+\overline T}\, , \qquad \cG_{S\overline S} = {1\over T+\overline T}.
\label{warped}\ee
Another instance of a non-flat geometry of the $S$-field  are the `axion stabiliser' terms~\cite{Carrasco:2015uma,Carrasco:2015pla} in the \K\, potential metric of the kind 
\be
S\overline S (T^i -\overline{ T}^i)^2 A(T^i, \overline{ T}^i)\, , \qquad \cG_{S\overline S} = (T^i - \overline{ T}^i)^2 A(T^i, \overline{ T}^i).
\label{stab}\ee
The inclusion of these terms {in some models}  is necessary for the stability of the inflationary trajectory. Finally, 
a new class of models where $\cG_{S\overline S} (T, \overline T)$ is a general Hermitian function was proposed in~\cite{McDonough:2016der}. A number of nice and interesting examples were studied, starting with $W= MS + W_0$ and shift symmetric canonical \K\, potentials, where also stability issues were studied, or with Poincar\'{e} half-plane geometries in \K\, potentials. 

An important feature of the  $\overline {D3}$ induced geometric inflation  models with one modulus is that the bisectional curvature   is non-vanishing, $R_{T \overline T S \overline S}\neq 0$ during inflation and at the exit, at the minimum of the potential. This is the consequence of the fact that the metric $\cG_{S\overline S}$  is not a product of a holomorphic $F(T)$ function times an  anti-holomorphic function $\overline F(\overline T)$. In the latter case it can be removed by a holomorphic change of the \K\, manifold coordinates $F(T) S \rightarrow S'$ which leads to a 
 a flat geometry of the nilpotent superfield. This case includes models with  canonical geometry for the nilpotent field, $\cG_{S\overline S} =1$ and some general superpotentials
$W = g(T^i) + f(T^i)S$. For these models the \K\, geometry of the nilpotent field $S$ is flat, and hence  $R_{i {\overline j} S \overline S}=0$.

A nice feature of our examples is that all of them during inflation, in case of a single modulus, have no tachyons without any assumption. At the minimum of the potential we do not have a general argument of stability, however, a priori these models allow a way to associate geometry with the good choices of the potentials which have a minimum at the exit from inflation. The same argument refers to multiple moduli models. A choice of the potentials is possible such that the desirable relations between moduli can be implemented as a requirement of the minimum of the potential, as a result we end up with single modulus models which have a stable inflationary trajectory.

Comparatively to other model building we used before, we have found various advantages, which we dubbed as a `model building paradise', based on a geometry of the $\overline{D3}$-brane and associated nilpotent multiplet interacting with moduli of the  Calabi-Yau manifolds.
In particular, we have a parameter of supersymmetry breaking independent of the Hubble parameter and the models are simple.

\section{Geometric Inflation Features}

\subsection{$\overline{D3}$-brane induced geometry}

We will  explain here that the most general \K\, invariant \K\,  function $\cG$ depending on multiple Calabi-Yau moduli and on a nilpotent multiplet $S$  can be reduced to  the form  we show in eq.~\rf{cG0}.
An equivalent form is to use 
\be
K (T^i, \overline{ T}^i; S, \overline S) = K_0(T^i, \overline{ T}^i) + S +\overline S + \cG_{S\overline S} (T^i, \overline{ T}^i)S\overline S\, , \qquad W= W_0 \,,
\label{hatK}\ee
where the gravitino mass, in general is given by the following expression:
\be
| m_{3/2} (T^i,\overline{ T}^i)|^2= e^{\cG_0(T^i,\overline{ T}^i)}=e^{K_0(T^i, \overline{ T}^i)} |W_0|^2 \,.
\ee
The linear terms in the \K~function and potential are directly related to spontaneous SUSY breaking and hence an integral aspect of our set-up. 

We will start with the observation~\cite{Kallosh:2015tea} that the most general supergravity theory with a number of unconstrained chiral multiplets $T^i$ and a single nilpotent superfield $S$ is given by  
 \begin{align}
{K}=&K_0(T^i,\overline{ T}^i)+K_S(T^i,\overline{ T}^i)S+\overline{K}_{\overline S} (T^i,\overline{ T}^i)\overline{S}+ \cG_{S\overline S}(T^i,\overline{ T}^i)S\overline{S} \,, \notag \\
 W =& g(T^i) + f(T^i)S,
 \end{align}
where $K_S,\overline{K}_{\overline S}$ and $\cG_{S \overline S}$ are non-holomorphic functions while $f$ and $g$ are holomorphic. These are the most general Taylor expansions of the \K~and superpotential due to the nilpotency condition $S^2 = 0$.
 
These general expressions can be simplified without loss of generality by a number of redefinitions. First of all, one can use a \K~transformation acting as
 \begin{align}
   W\to W\times\cF \,, \quad {K}\to K -\log |\cF|^2 \,,
 \end{align} 
to set $W=W_0$ by choosing $\cF = W_0/(g+fS)$. The resulting \K~potential is given by
 \begin{align}
 {K}'=&K'_0(T^i,\overline{ T}^i)+K'_S(T^i,\overline{ T}^i)S+\overline{K'}_S(T^i,\overline{ T}^i)\overline{S}+ \cG_{S\overline S}(T^i,\overline{ T}^i)S\overline{S},
 \end{align}
in terms of the redefined variables $K'_0=K_0 + 2 \log(|g| / |W_0|)$, and $K'_S=K_S+f/g$. In this frame, the supersymmetry breaking is set by $K'_S$ which we assume to be non-vanishing due to the nilpotency of $S$.
  
We can subsequently use the field redefinition $K'_SS = S$. Note that $K'_S$ is not holomorphic, and hence this field redefinition breaks the complex manifold structure. However, at least  in the bosonic part of the theory\footnote{The fermionic action will be affected by this change only in the part depending on the goldstino, a fermion in the   nilpotent multiplet, $\chi_S$. But in the unitary gauge, $\chi_S=0$, in which this fermion is absent, there will be no changes. In our models with $\cG_{T^i}=0$ in this gauge the gravitino decouples from the fermions in $T^i$ multiplets and the gauge with $\chi_S=0$ is simple.}
  this is not a problem for the following reason. The geometry spanned by the physical scalars is given by the K\"ahler manifold with a projection $S=0$, since the bosonic component of $S$ is a fermion bilinear, i.e.,
 \begin{align}
 ds^2= \cG_{i\overline j} dT^i d\overline{ T}^j|_{dS=S=0}.
 \end{align}
Therefore, the field redefinition $dS'=K'_SdS+\pd_iK'_SSdT^i+\pd_{\overline j}K'_SSd\overline{ T}^j$ does not change the K\"ahler manifold of physical scalars. It only affects the nilpotent part of the \K~potential, which now has a metric 
\be
\cG'_{S'\overline{S'}}= {\cG_{S\overline S}\over |K'_S|^2} \,.
\ee
This completes the argument that the most general supergravity theory can be brought to the  form \eqref{cG0} or \eqref{hatK} (omitting all primes),
when evaluated at $dS=S=0$.

In models satisfying our condition \rf{K} and with a positive CC, we can use the following form of the total potential where $V (T^i,\overline{ T}^i) $ vanishes at the minimum:
   \be
 \mathbf{ V} (T^i, \overline{ T}^i)=  V (T^i, \overline{ T}^i)  +\Lambda \,, \qquad 
\Lambda \equiv  |F_S|^2 -3 |W_0|^2 \,.
 \ee
This gives us an alternative form of the  $\overline {D3}$ geometry
\be
\cG_{S \overline S} (T^i, \overline{ T}^i)  =   {  |W_0|^2\over   |F_S|^2 + V (T^i, \overline{ T}^i)   }     \,,
\label{metric1}\ee
where the measure of supersymmetry breaking at $T^i= \overline{ T}^i$ due to the $\overline{D3}$-brane  is set by
\be
|\cG_S|^2 \equiv e^{\cG_0} \cG_S \cG^{S\overline S} \cG_{\overline S}= |F_S|^2 +V(T^i, \overline{ T}^i) \,.
\ee
Note that the above metric explicitly includes an independent Hubble, SUSY breaking and dark energy scale. 

In the absence of the nilpotent field, this model has a SUSY AdS solution with at least one flat direction amongst the $T^i$ moduli that will provide the inflaton. The inclusion of the $\overline{D3}$-brane yields the uplift term. When including a constant $\mu S$ term to the superpotential, or equivalently a constant metric $\cG_{S \overline S}$, this uplifts to a non-SUSY vacuum with arbitrary CC and a flat direction. The subsequent introduction of an inflationary profile can be performed either by means of a holomorphic function $f$ in the superpotential, or more generally by means of an moduli-dependent metric for the $S$-field, leading to the $\overline{D3}$-brane induced geometry~\eqref{metric}.

Also in more general models that do not satisfy \eqref{K}, we can reconstruct any desired potential $\mathbf{V}(T^i,\overline{ T}^i)$ starting from the \K\, function $\cG (T^i, \overline{ T}^i)$. In supergravity, the scalar potential and geometry are related as follows, assuming that  $\cG_S=1$:
 \begin{align}
    \mathbf{ V} (T^i, \overline{ T}^i)= e^ {{\cal   G}  (T^i,\overline{ T}^i)}  (\cG^{S \overline S} (T^i, \overline{ T}^i)  + \cG^{T^i \overline T^i}  {\cal   G} _{ T^i}  {\cal   G} _{\overline{ T}^i} - 3 ).
   \end{align} 
This relation is invertible with respect to $\cG_{S\overline{S}}$. In order to realize the desired potential $\mathbf{V}(T^i,\overline{ T}^i)$, we find the proper choice of $\cG_{S\overline S}$ is
   \begin{align}
\cG_{S \overline S} (T^i,\overline{ T}^i)  =   { e^ {{\cal   G} (T^i, \overline{ T}^i)} \over \mathbf{ V} (T^i,\overline{ T}^i)+3e^ {{\cal   G} (T^i, \overline{ T}^i)} - \cG^{T^i \overline T^i}  {\cal   G} _{ T^i}  {\cal   G} _{ \overline T^i} e^ {{\cal   G} (T^i, \overline{ T}^i)} }.
 \label{general}  \end{align}  
This geometry directly gives any phenomenologically favored potential.

\subsection{Curvature invariants}\label{curvature}

In case of one modulus $T$, this geometry is determined by two curvature invariants that will characterize the cosmological parameters. In addition to the full Ricci scalar, one can also define the Ricci scalar of the submanifold defined by $S=0$, as the only allowed coordinate redefinitions on this \K~geometry preserve the nilpotency condition. This will be referred to as the sectional curvature  and is given by \begin{align}
  R^{sec} & =-\cG^{T\overline T}\cG^{T\overline{T}}(\cG_{T\overline T  T\overline T}-\cG_{T T\overline T}\cG^{T\overline T}\cG_{T\overline T\overline T}) \,.
  \end{align}
The importance of this geometric quantity for inflationary model building has been stressed in various places. For example in the case of the hyperbolic disk relevant for $\alpha$-attractors, one has
 \begin{align}
    K= -3\alpha \ln (T+\overline T) \,, \quad R^{sec}=-{2\over 3\alpha} \,,
 \end{align}
where the latter is of course independent of the \K~frame.

The new ingredient in the $\overline {D3}$ induced geometric inflation models is the second curvature invariant, corresponding to the  bisectional curvature along the $S=0$ plane:
 \begin{align}
  R^{bisec} & = - R_{T\overline T S\overline S} \cG^{T \overline T} \cG^{S \overline S} = \frac{\cG^{T \overline T}(V_{T \overline T}(F_S^2+V)-V_TV_{\overline T})}{(F_S^2+V)^2} \,.
 \label{bisec}
 \end{align}
 During inflation at $V\gg |F_S|^2$, it is proportional to slow roll parameters
\begin{align}
R^{bisec} |_{infl}\approx   \cG^{T\overline T} \Big ({V_{T\overline T}\over V}-{V_TV_{\overline T} \over V^2}\Big)=  \eta-2\epsilon \,.
\label{bisec1}\end{align}
In contrast, at the minimum of the potential,
\begin{align}
R^{bisec}|_{min}=-R_{T\overline T S\overline S}\cG^{T\overline T}\cG^{S\overline S} =\frac{\cG^{T\overline T}V_{T\overline T}}{F_S^2} >0 \,.
\label{bisec2}\end{align}
It therefore sets the scale for the sum of masses of both $T$-components, and stability requires a positive value for the bisectional curvature.

\subsection{Stability analysis}\label{sinflaton}

For a model with a single inflaton superfield model, we find that the supersymmetric scalar partner of inflaton (the so-called sinflaton) is always stabilized at its origin as shown below.

The general formula for the non-holomorphic masses of the scalar fields is given in the notation of~\cite{Ferrara:2016ntj} by 
\begin{align}
m_{ i\overline j}^2=e^{\cal G}[\cG_{i\overline{j}}\Big (1+\frac{\bf V}{|m_{3/2}|^2}\Big )-{\cal G}_i {\cal G}_{\overline j}+({\cal G}_{i\alpha}+{\cal G}_i {\cal G}_{\alpha})\cG^{\alpha\overline\beta}({\cal G}_{\overline{\beta}\overline{j}}+{\cal G}_{\overline\beta}{\cal G}_{\overline j})-R_{i\overline j \alpha\overline \beta} {\cal G}^\alpha {\cal G}^{\overline\beta}],
\end{align}
where ${\cal G}^{\alpha}\equiv \cG^{\alpha\overline\beta}{\cal G}_{\overline\beta}$, $\alpha = (S, T^i)$ and $i=1,\ldots,N$ . Under the assumption~\eqref{K} for the physical scalar fields, this simplifies to
\begin{align}
m_{ i\overline j}^2=e^{\cal G}[\cG_{i\overline{j}}\Big (1+\frac{\bf V}{|m_{3/2}|^2}\Big ) + {\cal G}_{i\alpha} \cG^{\alpha\overline\beta} {\cal G}_{\overline{\beta}\overline{j}} - R_{i\overline j S\overline S} {\cal G}^S {\cal G}^{\overline S}],
\end{align}
Tracing this formula yields the average mass: 
\begin{align}
m^2_{ave} =  \frac1N \cG^{i \overline j}m^2_{i\overline j}=   \frac{1}{N}e^\cG\left(N\Big (1+\frac{ \bf V}{|m_{3/2}|^2}\Big ) +\cG_{i\alpha}\cG_{\overline j \overline\beta}\cG^{\alpha\overline\beta}\cG^{i\overline j} -  \cG^{i\overline j}R_{i\overline j S\overline S} {\cal G}^S {\cal G}^{\overline S} \right).
\end{align}
In particular, for $N=1$, this expression can be reduced to
\begin{align}
m^2_{ave}=&{\bf V}+m_{3/2}^2 + m_{3/2}^2 (|\cG_{TT}|^2(\cG^{T\overline T})^2-\cG^{T \overline T}R_{T\overline T S\overline S}\cG^{S}\cG^{\overline{S}})\nn\\
=&{\bf V}+m_{3/2}^2 + m_{3/2}^2 |\cG_{TT}|^2(\cG^{T\overline T})^2+R^{bisec}(|F_S|^2+V) \,,
\end{align}
emphasizing the importance of the bisectional curvature.

During inflation, the inflaton mass is very small, and the first 3 terms are positive. The last term is given by the linear combination of slow-roll parameters \eqref{bisec1}. Using the experimental values of $n_s$ and $r$ it comes out negative, but is always smaller than the first two positive contributions thanks to the slow-roll suppression. Thus, during inflation, we have shown that the sinflaton direction in a single superfield model is always stable, or equivalently our assumption $T=\overline T$ is satisfied automatically.

Apart from the inflationary era, we discuss the minimum of our model. The nilpotent superfield is well defined only if $\cG_S\neq0$ and $\cG_{S\overline S}\neq0$. Due to the absence of a propagating scalar in $S$, the stability requirement is equivalent to the condition that the propagating scalars have stable vacua at $\cG_S\neq0$ and also $\cG_{S\overline S}\neq0$. Then, we need to require positive masses for the scalar fields at the minimum. The general minimization condition of the scalar potential is
\begin{align}
{\bf V}_i=\cG_i{\bf V}+e^{\cG}(\nabla_i\cG_\alpha \cG^{\alpha\overline \beta}\cG_{\overline \beta}+\cG_i)=0.
\end{align}  
Since ${\bf V}=\Lambda\sim 0$ at the minimum, we obtain the condition $\nabla_i\cG_\alpha \cG^{\alpha\overline \beta}\cG_{\overline \beta}+\cG_i=0$. For $\cG_i=0$, the condition is equivalent to $\nabla_i\cG_S=0$. Then, the mass matrix at the minimum is simply given by
\begin{align}
m_{ i\overline j}^2=e^{\cal G}[\cG_{i\overline{j}}+{\cal G}_{ij}\cG^{j\overline k}{\cal G}_{\overline{k}\overline{j}}+R_{i\overline j S\overline S}(\cG^{S\overline S})^2].
\end{align}
Assuming ${\cal G}_{ij}\cG^{j\overline k}{\cal G}_{\overline{k}\overline{j}}=\mathcal O(1)$ and $R_{i\overline j S\overline S}=0$, the averaged mass becomes
\begin{align}
m_{ ave}^2=\mathcal{O}(m_{3/2}^2).
\end{align}
Therefore, to disentangle the scalar mass and the SUSY breaking scale, we need to introduce large ${\cal G}_{ij}\cG^{j\overline k}{\cal G}_{\overline{k}\overline{j}}$ or $R_{i\overline j S\overline S}(\cG^{S\overline S})^2$. Moreover, the 
 the scale of the averaged mass does not tell us the mass of each scalar and their positivity, and therefore, we need to discuss the stability at the minimum for each case.

With our choice of $\cG_{S\overline S}$ in the single-modulus $N=1$ case, the averaged mass becomes
\begin{align}
m_{ ave}^2=e^{\cal G}(1+{\cal G}_{TT}\cG^{T\overline T}{\cal G}_{\overline{T}\overline{T}}) +\cG^{T\overline{T}}V_{T\overline T}.\label{avm}
\end{align}
The last term comes from the bisectional curvature and it is not necessarily related to the SUSY breaking scale. Thus, with a proper choice of $V$, the SUSY breaking and the mass of the inflation sector can be disentangled.

\section {Model Building Paradise}

Our main goal here is to give example of geometric models of inflation which are defined by a geometry of the $\overline{D3}$-brane in the CY bulk geometry. For this purpose it is natural to use logarithmic \K\, potentials for the moduli fields $T^i$ of the kind $\ln (T^i+\overline T^i)$. However, once we use nilpotent superfield geometry  as a tool in model building, we find that the shift symmetric \K\, potentials for the moduli fields $\Phi^i$ are also particularly efficient. We will start therefore  with the  model of polynomial inflation with the  \K\, potential $-{1\over 2}(\Phi - \overline \Phi)^{2}$.

\subsection{Polynomial inflation}

Inflation-related Planck data~\cite{Planck:2015xua} describing Gaussian adiabatic perturbations consist of 3 main parameters: the amplitude of the perturbations $A_{s}$, the  spectral index $n_{s}$ and the tensor to scalar ratio $r$.  According to~\cite{Destri:2007pv,Nakayama:2013jka}, one can properly describe   any  set of these   parameters in the context of the 3-parameter polynomial inflationary models with the 
potential 
\be
V(\phi) = {m^{2}\phi^{2}\over 2}(1 + a \phi+ b\phi^{2}). \label{pol}
\ee
One could try to implement the models with such potentials in supergravity~\cite{Nakayama:2013jka}, using the general approach developed in~\cite{Kawasaki:2000yn,Kallosh:2010ug,Kallosh:2010xz}, but the resulting potentials can reproduce the potential~\rf{pol} only approximately, see a discussion of this issue in~\cite{Kallosh:2014xwa}. Meanwhile, as we will see now, the potential~\rf{pol} can be easily obtained in the context of the new geometric approach discussed in our paper.

\begin{figure}[t!]
\begin{center}
\includegraphics[scale=0.65]{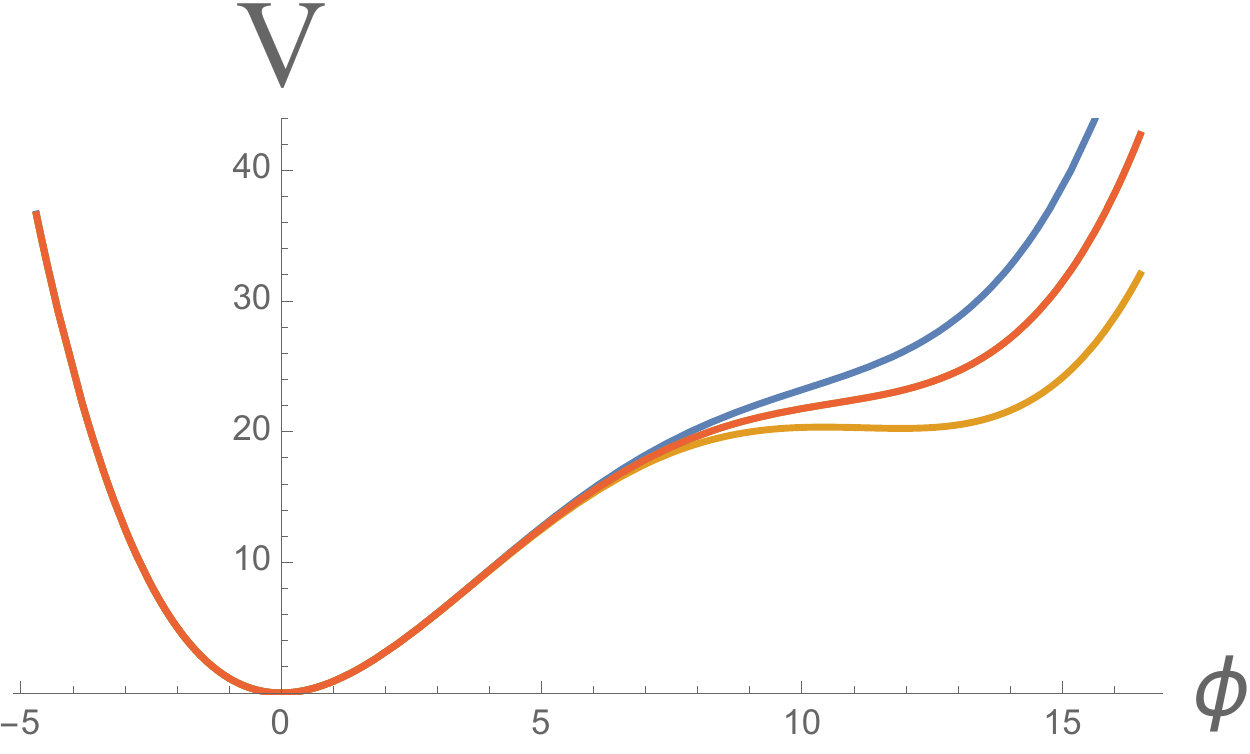}
\end{center}
\caption{\footnotesize The potential $V(\phi) = {m^{2}\phi^{2}\over 2}\,\bigl(1-a\phi +a^{2}b\,\phi^{2}\bigr)$ for $a = 0.12$ and $b = 0.30$ (upper curve), $b = 0.29$ (middle), and $b = 0.28$ (lower curve). The potential is shown in units of $m^{2}$, with $\phi$ in Planck units.  For  $b = 0.29$ (the middle curve), at the moment corresponding to $N=  58$ e-folding from the end of inflation one has $n_{s}= 0.965$ and $r = 0.012$, perfectly matching the Planck data. }
\label{chi}
\end{figure}

We will consider the \K\ function
\begin{align}
{\cal G}=&\log W_0^2-{1\over 2}(\Phi - \overline \Phi)^{2} + (S+\overline S)+g_{S\overline S}S\overline{S},  \quad
g^{S\overline S}=\frac{1}{W_0^2}\left(|F_S|^2+V(\Phi, \overline \Phi) \right). \label{shift}
\end{align}
Here the part of the potential vanishing at the minimum is 
\be 
V(\Phi, \overline \Phi) = {m^{2}\over 4}(\Phi + \overline \Phi)^{2} \Bigl(1 - {a\over \sqrt 2}(\Phi + \overline \Phi)\bigl(1+{a\, b\over \sqrt 2} (\Phi + \overline \Phi)\bigr)\Bigr)  .
\label{p}\ee
We represent the field $\Phi$ in terms of its canonically normalized components, $\Phi = {1\over \sqrt 2} (\phi + i \chi)$. One can show that the potential of these fields is stable at $\chi = 0$, and the inflaton fields $\phi$ has the desirable potential 
\be
{\bf V}(\phi) = {m^{2}\phi^{2}\over 2}\,\Bigl(1-a\phi(1 +b\, a\, \phi )\Bigr) +\Lambda,
\ee
where $\Lambda = |F_S|^2- 3|W_{0}|^{2}$ is the vacuum energy/cosmological constant at the minimum of the potential, and the gravitino mass at the minimum is equal to $m_{3/2} = W_{0}$.  The potential for $\Lambda = 0$ is shown in Fig. \ref{chi}.

As we already mentioned, inflation-related Planck data \cite{Planck:2015xua} consist of three main parameters, $A_{s}$, $n_{s}$ and $r$.   The value of $A_{s}$  can be easily tuned by a proper choice of $M$. The parameters $a$ and $b$ are responsible for $n_{s}$ and $r$. For example, for $a = 0.12$ and $b = 0.29$, the perturbations generated at the moment corresponding to $N=  58$ e-folding from the end of inflation have $n_{s}= 0.965$ and $r = 0.012$, perfectly matching the Planck data~\cite{Planck:2015xua}. 

Thus we found the desirable polynomial potential, and much more: we have full flexibility to describe arbitrary cosmological constant and SUSY breaking in this simple model.  Finally, inflation in this model may begin close to the Planck density, which easily solves the problem of initial conditions for inflation, as explained in~\cite{Linde:2005ht}. 

\subsection{T-models}

Moving on to a hyperbolic instead of a flat geometry for the scalar manifold, the \K\, function in disk variables can be written as
\be
\cG=\ln W_0^2 -{3\alpha \over 2} \log   {(1-Z\overline Z)^2\over (1-Z^2) (1-\overline Z^2)}  +S + \overline S + {W_0^2 \over |F_S|^2+ m^2 Z\overline Z} S \overline S.
\ee
Note that this employs a \K~frame that has a manifest inflaton shift symmetry  \cite{Carrasco:2015uma}. 
One can check that $\cG_Z=0$ and $\cG_S=1$, i.e. the theory has all required properties.  

\begin{figure}[t!]
\begin{center}
\includegraphics[scale=0.6]{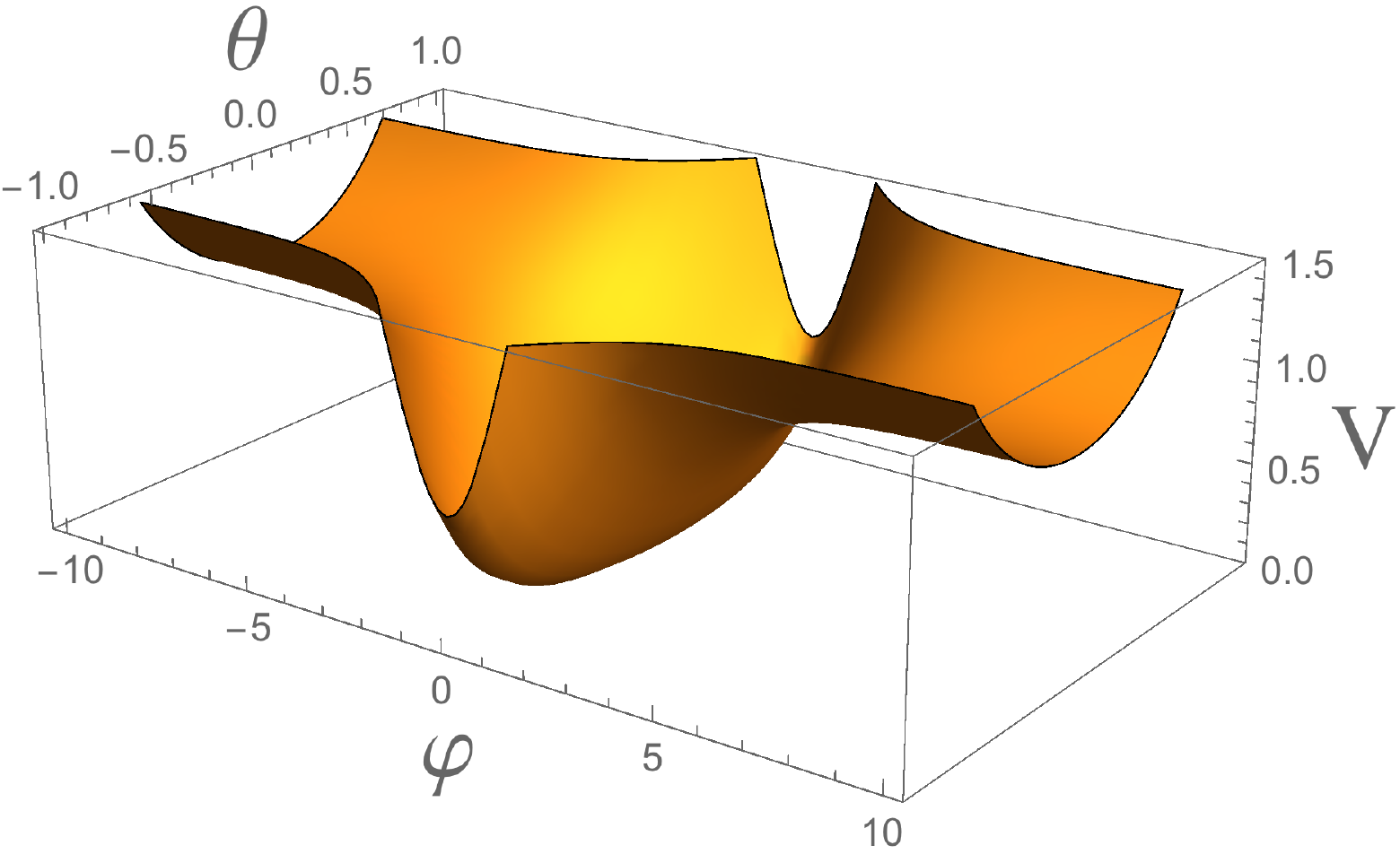}
\end{center}
\caption{\footnotesize Basic T-model with  $\alpha = 1$. The height of the potential here and in other figures is in units $m^{2}$ and the values of the fields are in Planck mass units.}
\label{2T}
\end{figure}

The canonical inflaton $\vp$ is defined by relation $Z= \tanh {\vp\over \sqrt{6\alpha}}$. The inflaton potential  is  
\be
 { \bf V} |_{Z=\overline Z}= \Lambda + m^2 \tanh^{2} {\vp\over \sqrt{6\alpha}} \ ,
\ee
where $\Lambda =|F_S|^2- 3W_0^2$. The axion mass along the inflaton trajectory for $\Lambda = 0$ is 
\begin{align}
m_\theta^2=2(m^2+2W_0^2)-m^2\left(\Bigl(2-{2\over 3\alpha} \Bigr)\cosh ^2 \frac{\varphi}{\sqrt6\alpha}+{1\over 3 \alpha}\right)\left(\cosh\frac{\varphi}{\sqrt6\alpha}\right)^{-4}.
\end{align}
As expected from the observation in Sec.~\ref{sinflaton}, the mass of the axion $\theta$ is positive during inflation: $m_\theta^2=2(m^2+2W_0^2) > 2 V \sim 6 H^{2}$ for $\vp \gg  \sqrt{6\alpha}$. This means that the field $\theta$  is strongly stabilized and its perturbations are not generated during inflation.  Moreover, the stability condition is satisfied along the full inflaton trajectory for all $\phi$. In particular, the masses of the fields $\vp$ and $\theta$ at the minimum of the potential at $\vp = \theta = 0$ are given by
\be
m^{2}_{\phi} = {m^{2}\over 3\alpha}\ , \qquad m^{2}_{\theta} = {m^{2}\over 3\alpha} +4W_{0}^{2} \ .
\ee
These results are illustrated by Fig.~\ref{2T}, which shows the  potential $V(\vp,\theta)$ in the limit $W_{0}^{2} = m_{3/2}^{2} \ll m^{2}$ for the particular case $\alpha = 1$.

If we use a more general function
\be
\cG=\ln W_0^2 -{3\alpha \over 2} \log   {(1-Z\overline Z)^2\over (1-Z^2) (1-\overline Z^2)}  +S + \overline S + {W_0^2 \over |F_S|^2+ f(Z\overline Z)} S \overline S,
\ee
and the potential is
\be
{\bf V}  |_{Z=\overline Z}= F_S^2- 3W_0^2 + f(\tanh^2 {\varphi\over \sqrt{6\alpha}}).
\ee
Then, the axion mass becomes
\begin{align}
m_{\theta}^2=&4W_0^2+2f+\frac{{\rm cosh}\sqrt{\frac{2\phi}{3\alpha}}\, {\rm sech}^4\frac{\phi}{\sqrt{6\alpha}}\,f'}{3\alpha},
\end{align}
where the prime denotes the derivative with respect to the argument $\tanh^2 {\varphi\over \sqrt{6\alpha}}$.
The last term becomes $O(\sqrt{\epsilon})H^2$ whereas the second term is $6H^2$ and is much larger than the last term. Therefore, the axion mass is positive as we expected from the general discussion in Sec.~\ref{sinflaton}. The minimum is $\phi=0$ and the mass of the inflaton and axion at the minimum are 
\begin{align}
m_{\phi}^2=&\frac{f'(0)}{3\alpha}, \quad 
m_{\theta}^2 = 4W_0^2+\frac{f'(0)}{3\alpha},
\end{align} 
where we have used $f(0)=0$, which is our general assumption on $V$. One can check that this coincides with the general formula \eqref{avm} from the previous section.

\subsection{E-models}

A simple case using half-plane variables
is
\be
\cG=\ln W_0^2 -{3\alpha \over 2} \log   {(T+ \overline T)^2\over 4T \overline T}  +S + \overline S + {W_0^2 \over |F_S|^2+ m^2 \bigl(1- {T+ \overline T\over 2}\bigr)^{2}} S \overline S. \label{disk}
\ee
The trajectory is stable at $T=\overline T$.  The canonical inflaton $\vp$ is defined by $T= e^{- \sqrt{2\over 3\alpha}\varphi}$. The inflaton potential  is  
\be
{\bf V}_\Lambda |_{T=\overline T}=\Lambda + m^2 \Bigl(1-e^{ -\sqrt{2\over 3\alpha}\varphi}\Bigr)^2. \label{Emod}
\ee
One can check that $\cG_T=0$ and $\cG_S|_{\min} = 1\neq0$, i. e. the theory has all required properties. The axion mass squared during inflation is 
\begin{align}
m_{a}^2=2m^2\left(1-e^{-\sqrt{\frac{2}{3\alpha}}\varphi}\right)^2+4W_0^2.
\end{align}
It is positive definite during and after inflation. Note that at the minimum $\varphi=0$, the axion mass squared becomes $4W_0^2=4m_{3/2}^2$.
\begin{figure}[t!]
\begin{center}
\includegraphics[scale=0.6]{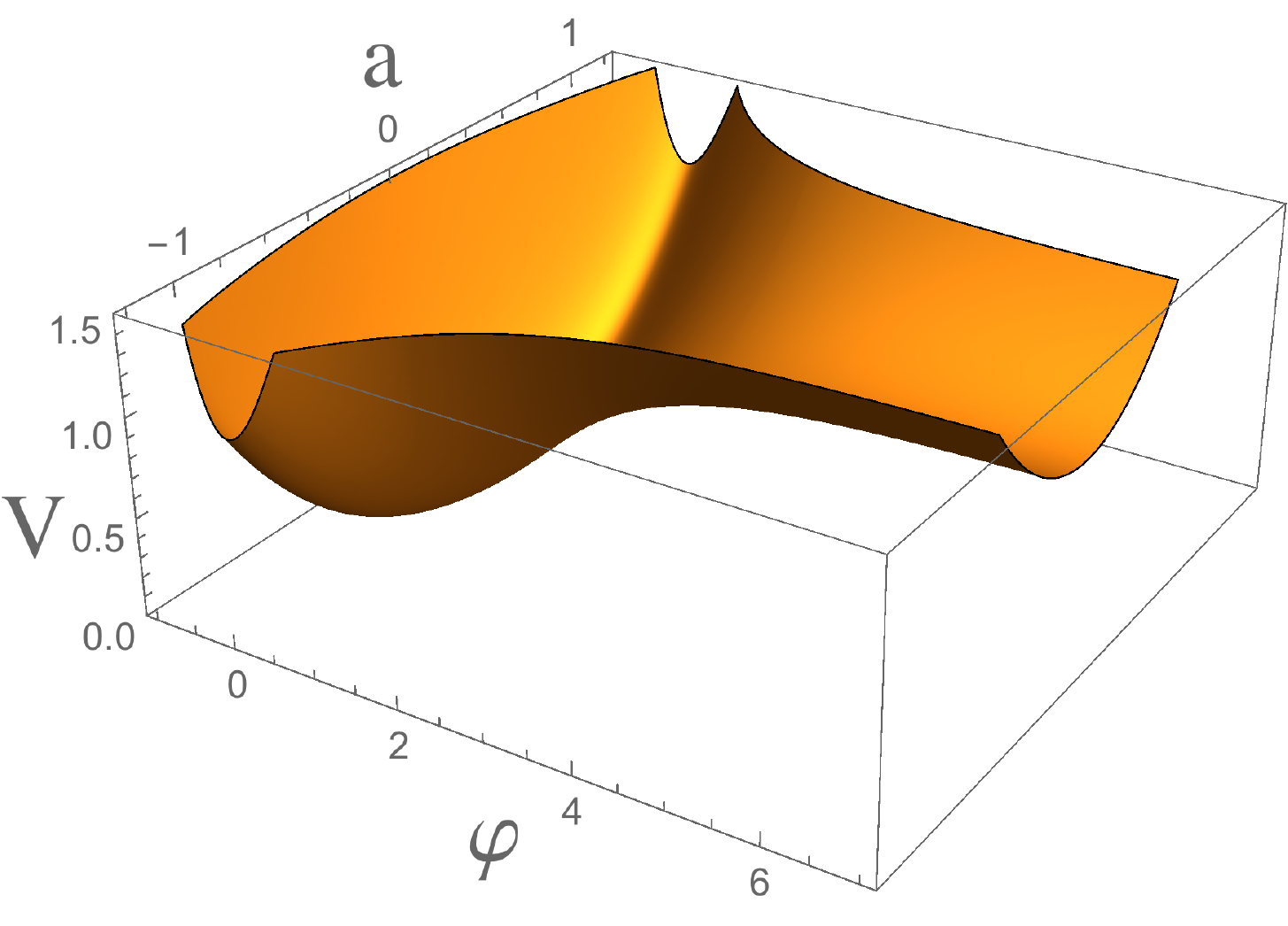}
\end{center}
\caption{\footnotesize Basic E-model with  $\alpha = 1/3$. }
\label{2T2}
\end{figure}

If we take a more general function
\be
\cG=\ln W_0^2 -{3\alpha \over 2} \log   {(T+ \overline T)^2\over 4T \overline T}  +S + \overline S + {W_0^2 \over |F_S|^2+ m^2 f(T+ \overline T)} S \overline S,
\ee
and the potential is
\be
{\bf V} |_{Z=\overline Z}= |F_S|^2- 3W_0^2 + M^2 f\bigl(2e^{ \sqrt{2\over 3\alpha}\varphi}\bigr) \ .
\ee
In this case, the mass of the axion is given by
\begin{align}
m_{a}^2=4W_0^2+2f\bigl(2e^{ \sqrt{2\over 3\alpha}\varphi}\bigr)=4W_0^2+6H^2,
\end{align}
which is positive definite and consistent with our general argument in Sec.~\ref{sinflaton}.

\subsection{Two-disk merger models}

\subsubsection{E-model}

\begin{figure}[t!]
\begin{center}
\includegraphics[scale=0.63]{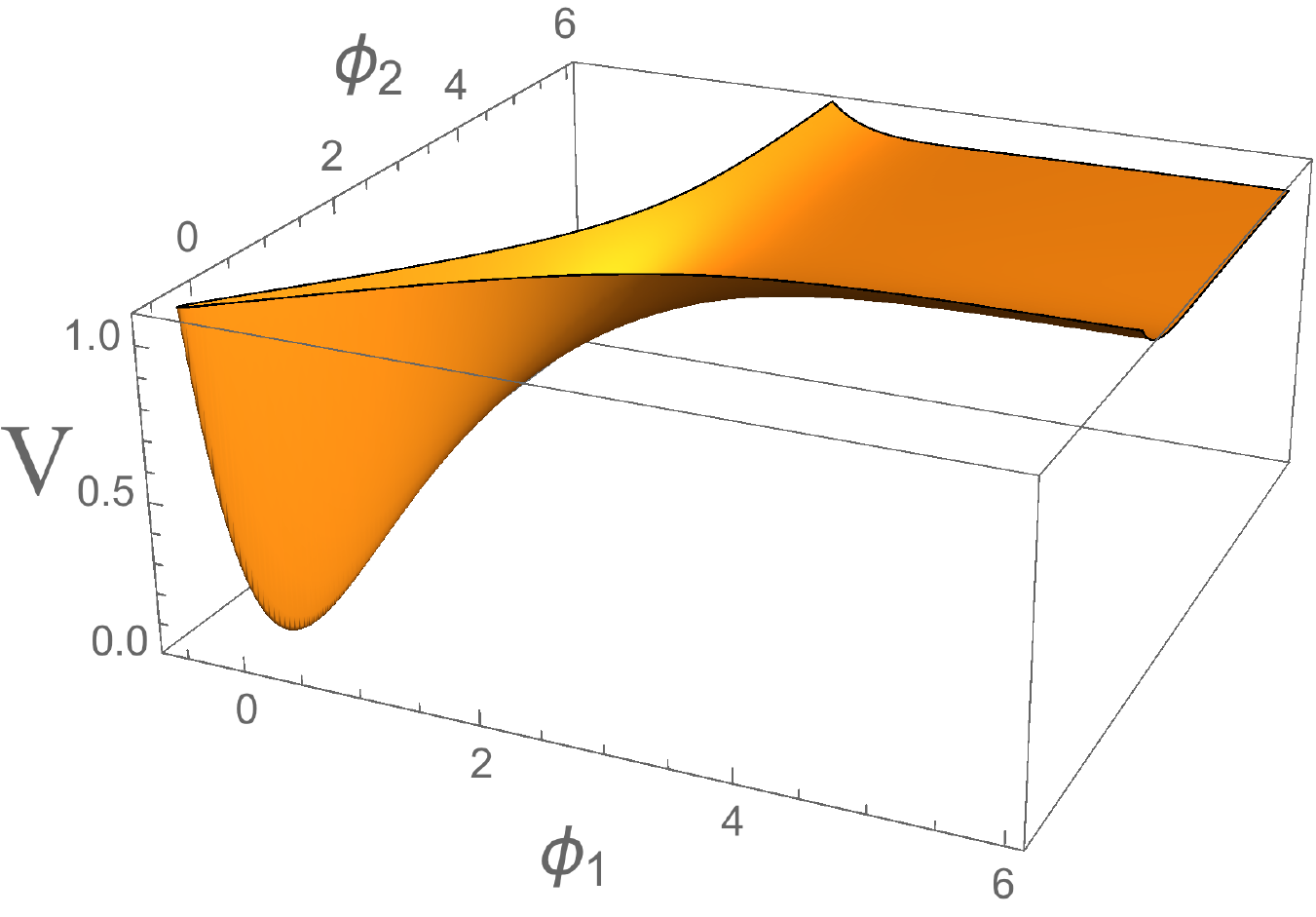}
\end{center}
\caption{\footnotesize Merger of two disks with $\alpha =1/3$ creates the inflaton potential with $\alpha = 2/3$.  Here we considered an example with $M = 6\, m$.} \label{2E}
\end{figure}

Here we consider the model with two half-planes $T_{1,2}$ and $3\alpha_i=1$ for $i=1,2$. As the previous work~\cite{Kallosh:2017ced} where the merger of different attractors was discussed (albeit in disk coordinates), we dynamically realize the inflationary trajectory where two half-plane moduli directions merge during last 50-60 e-foldings. Instead of the use of the superpotential for stabilization~\cite{Kallosh:2017ced}, we use the geometry,
\begin{align}
{\cal G}=&\log W_0^2-\frac12\sum_{i=1}^2\log \left(\frac{(T_{i}+\overline{T}_i)^2}{4T_{i}\overline{T}_{i}}\right)+S+\overline{S}+g_{S\overline S}S\overline S,\\
g^{S\overline S}=&\frac{1}{W_0^2}\left(|F_S|^2+m^2\bigl(1-\frac{1}{4}(T_{1}+\overline{T}_{1}+T_{2}+\overline{T}_{2})\bigr)^2+\frac{1}{4}M^2(T_{1}+\overline{T}_{1}-T_{2}-\overline{T}_{2})^2\right).
\end{align}
Then the scalar potential is 
\begin{align}\label{stabE}
{\bf V}=\Lambda +m^2\bigl(1-\frac{1}{4}(T_{1}+\overline{T}_{1}+T_{2}+\overline{T}_{2})\bigr)^2+\frac{1}{4}M^2(T_{1}+\overline{T}_{1}-T_{2}-\overline{T}_{2})^2.
\end{align}
The last term in \rf{stabE} leads to the merger of inflationary trajectories of $T^i$ as shown in Fig.~\ref{2E}.
We represent field $T_{i}$ as  $T_{i}=e^{-\sqrt{2}\phi_i}(1+\sqrt{2}\theta_i)$, where  $\phi_i$ are canonical, and $\theta_i$ are canonical in the small $\theta_i$ limit. The inflaton direction on merger trajectory is $\varphi=\frac{1}{\sqrt2}(\phi_1+\phi_2)$ and the orthogonal direction is $\chi=\frac{1}{\sqrt{2}}(\phi_1-\phi_2)$. During inflation with $\varphi=\frac{1}{\sqrt2}(\phi_1+\phi_2)$ the potential of the canonically normalized inflaton field $\varphi$ is
\begin{align}
{\bf V}(\vp)=\Lambda +m^2\left(1-e^{-\vp} \right)^2
\end{align}
corresponding to $3\alpha =2$.

In models with multiple fields, the stability of the axionic direction is not guaranteed by the discussion in Sec.~3.1 and we need to discuss the stability of the trajectory for each case. For the current model, the axionic directions are stabilized with masses
\begin{align}
m_{a_1}^2=m_{a_2}^2=2(m^2(1-e^{-\varphi})^2+2W_0^2)=6H^2+4W_0^2,
\end{align}
where $H^2=\frac13 V=\frac13m^2(1-e^{-\varphi})^2$. As is the case of the previous work~\cite{Kallosh:2017ced}, the direction $\chi=\frac{1}{\sqrt{2}}(\phi_1-\phi_2)$ acquires a light or tachyonic mass for sufficiently large $\phi$:  the mass of $\chi$ is given by
\begin{align}
m_\chi^2=2e^{-2\varphi}(4M^2-m^{2}(e^{\varphi}-1)).
\end{align}
As explained in~\cite{Kallosh:2017ced}, this simply means that the exponentially flat and long dS plateau in the upper right corner of Fig.~\ref{2E} is slightly curved, and the fields tend to move towards its boundaries. Then they slide along these boundaries towards the point where these boundaries merge and the diagonal deep gorge is formed, as shown at the center of  Fig.~\ref{2E}. After that, all fields become stable along the inflationary trajectory with $\phi_1=\phi_2=\frac{1}{\sqrt2}\varphi$ and the inflaton potential coincides with the E-model potential \rf{Emod}. The field value of $\varphi$ at the last $N$ e-folding is given by $\varphi_N=\log(4N)$. The condition that the merger trajectory is stable for last $N$ e-foldings is $M^2>\frac{m^2N}{2}$.

At the minimum, $\cG_S=1$, the metric is $\cG_{S\overline S}=\frac{W_0^2}{|F_S|^2}\sim\frac{1}{3}$, and the SUSY breaking is realized  with $m_{3/2} = W_{0}$. Thus, this model generalizes the E-model disk merger described in~\cite{Kallosh:2017ced}, but now one can have arbitrary values of the cosmological constant $\Lambda$ and the gravitino mass.

\subsubsection{T-model}

\begin{figure}[t!]
\begin{center}
\includegraphics[scale=0.6]{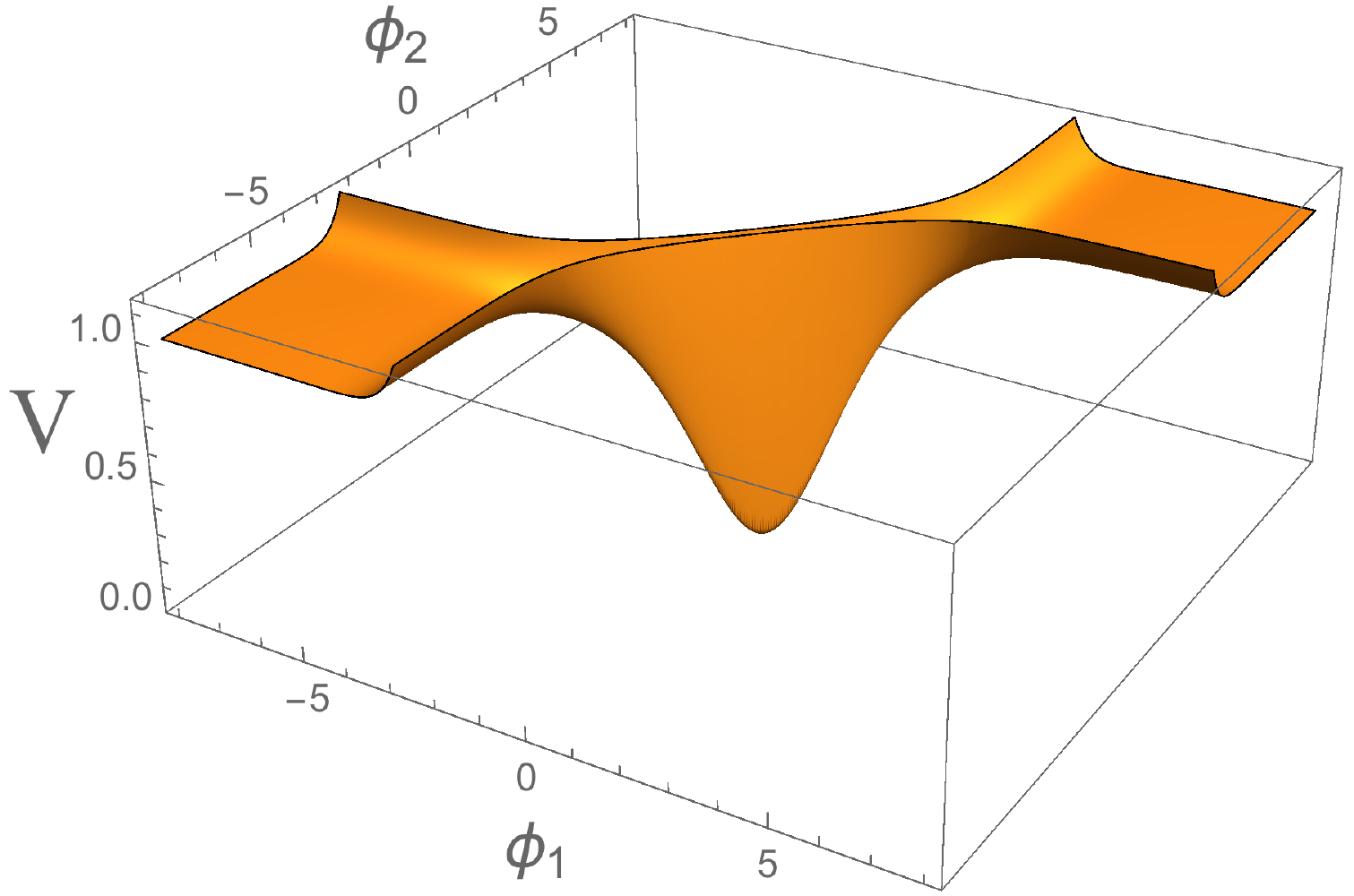}
\end{center}
\caption{\footnotesize Merger of two disks with  $\alpha =1/3$ creates the inflaton T-model potential with $\alpha = 2/3$. In this figure we show the potential with  $M = 10\,  m$.}\label{2T3}
\end{figure}

The disk merger model is also possible for T-models. We consider the following system, 
\begin{align}
{\cal G}=&\log W_0^2-\frac12\sum_{i=1}^2\log\frac{(1-Z_i\overline{Z}_i)^2}{(1-Z_i^2)(1-\overline{Z}_i^2)}+S+\overline{S}+g_{S\overline S}S\overline{S},\\
g^{S\overline S}=&\frac{1}{W_0^2}\left(|F_S|^2+\frac{m^2}{2}(|Z_1|^2+|Z_2|^2)+\frac{M^2}{4}\left((Z_1+\overline{Z}_1)-(Z_2+\overline{Z}_2)\right)^2 \right).
\end{align}
The scalar potential is \begin{align}
{\bf V}= \Lambda+ \frac{m^2}{2}(|Z_1|^2+|Z_2|^2)+\frac{M^2}{4}\left((Z_1+\overline{Z}_1)-(Z_2+\overline{Z}_2)\right)^2,
\end{align}
and the last term gives the dynamical constraint $\phi_1=\phi_2$ where  we have defined canonical fields as $Z_i=\tanh \frac{\phi_i+{\rm i}\theta_i}{\sqrt{2}}$. During inflation with  $\phi_1=\phi_2=\frac{1}{\sqrt2}\varphi$ the potential is
\be\label{Tmod}
{\bf V}(\vp)= \Lambda+ {m^2} \tanh^2 {\varphi \over 2}
\ee
corresponding to $3\alpha =2$. The scalar potential is shown in Fig.~\ref{2T3}.

Turning to stability, the mass eigenvalues of axionic directions on inflationary trajectory $\phi_1=\phi_2=\frac{1}{\sqrt2}\varphi$ are given by
\begin{align}
m_1^2=m_2^2=4W_0^2+\frac{2m^2(\cosh^{2}\varphi+\cosh\varphi-1)}{(\cosh\varphi+1)^2}.
\end{align}
The masses are positive. At the minimum, $m_i^2=\frac{m^2}{2}+4W_0^2$. Instead, the mass of the $\chi$ direction is 
\begin{align}
m_{\chi}^2=\Bigl(m^2+2M^2-\frac{m^2}{2}\cosh\varphi\Bigr)\ \cosh^{{-4}}\frac{\varphi}{2} .
\end{align}
As in the E-model discussion, for the very large values of the inflaton field, such that  $m^2\cosh\phi > 4M^2$, the field $\chi$ is tachyonic. In order for this instability to take place outside of the observable window of $N$ e-folds, one has again has to impose the condition $M^2>\frac{m^2N}{2}$.
 
As in the previous section, at the minimum, $\cG_S=1$, the metric is $\cG_{S\overline S}=\frac{W_0^2}{|F_S|^2}\sim\frac{1}{3}$, and the SUSY breaking is realized  with $m_{3/2} = W_{0}$. Thus, this model generalizes the T-model disk merger described in~\cite{Kallosh:2017ced}, but now one can have arbitrary values of the cosmological constant $\Lambda$ and the gravitino mass.

\subsubsection{Cascade inflation}

\begin{figure}[t!]
\begin{center}
\includegraphics[scale=0.7]{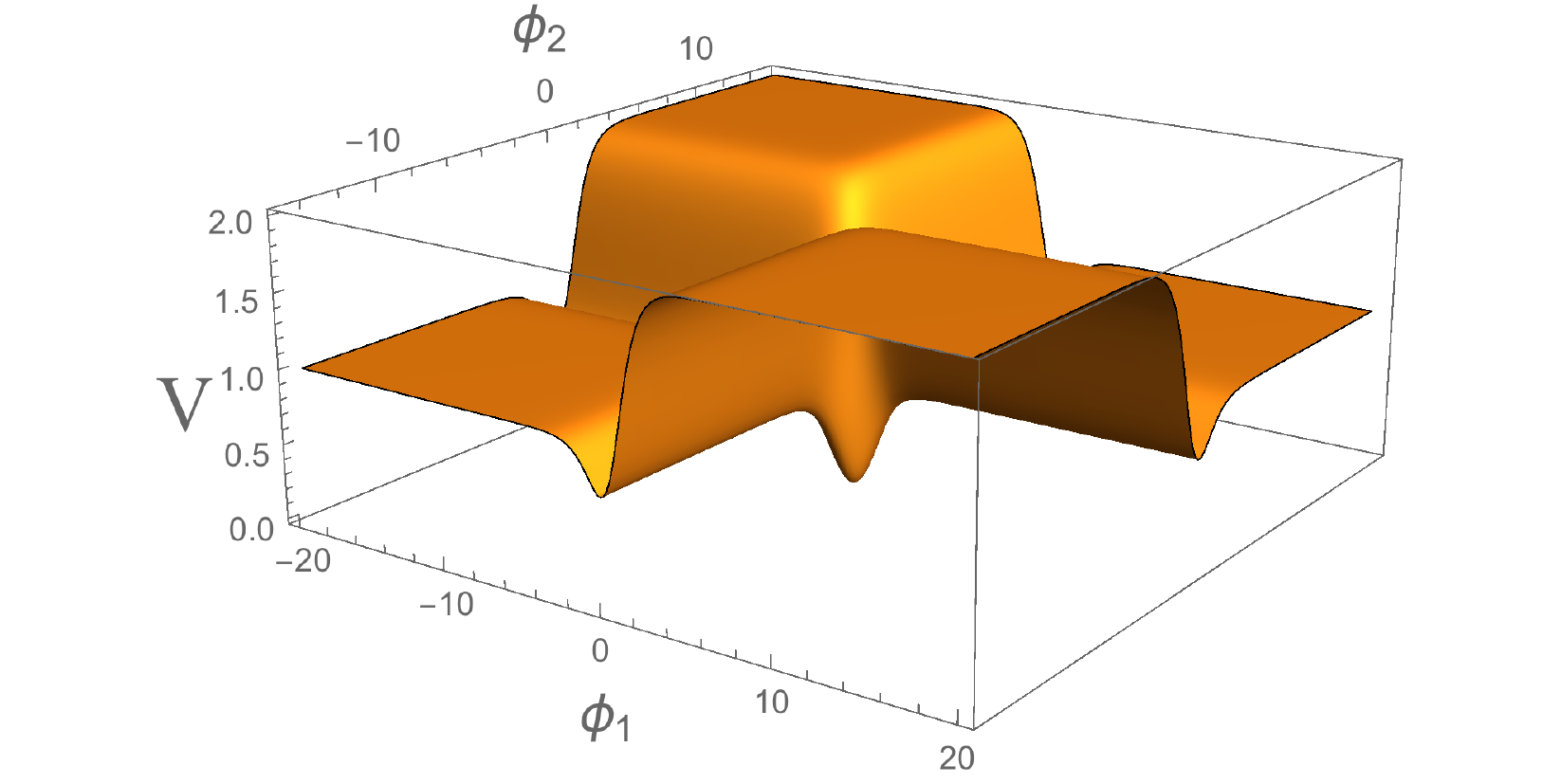}
\end{center}
\caption{\footnotesize The potential of the two disks with $\alpha =1/3$ for $m = M$. }
\label{2Tup}
\end{figure}

The two-disk merger (the fusion of two different attractors) is not the only interesting feature of the two-disk model studied above. Fig.~\ref{2T3} shows only the lower part of the potential, which is sufficient to illustrate the effect of the disk merger. However, the upper part of the potential tells us an equally interesting story. To explain it, we will show the potential including its upper part, for a toy model with $m = M$, see Fig.~\ref{2Tup}.
One can easily recognize the minimum of the potential, near which one may have inflation with $\alpha = 2/3$ for the models with $M\gg m$. However, another important part of the potential is the existence of 4 different dS plateaus. The lower ones have the height  $m^{2}$, one can see them also in Fig.~\ref{2T3}. The upper ones have the height $m^{2}+M^{2}$. They exist even in the absence of the disk interactions, for $M=0$, in which case the height of each plateau is equal to $m^2 $. 

The existence of these plateaus follows from the general expression for the potential of the fields $\phi_1$ and $\phi_2$ in that model:
\be\label{Tmod2}
{\bf V}(\vp)= \Lambda+ {m^2\over 2} \left(\tanh^2 {\phi_1 \over \sqrt 2} + \tanh^2 {\phi_2 \over \sqrt 2}\right) +{M^2\over 4}\left(\tanh {\phi_1 \over \sqrt 2} - \tanh {\phi_2 \over \sqrt 2}\right)^2 \ .
\ee
Inflation may begin at the upper plateau, with $\phi_1\gg 1$ and $-\phi_2 \gg 1$, or with $-\phi_1\gg 1$ and $\phi_2 \gg 1$. Then the field falls down to one of the lower plateaus, from which it  moves towards the narrow gorge  along $\phi_1 = \phi_2$  in the potential shown in Fig.~\ref{2T3}, and eventually falls to the minimum at $\phi_{1}= \phi_{2}= 0$. One may call this multi-stage process {\it a cascade inflation.} For $M^2>\frac{60 m^2}{2}$, all observational consequences of this regime are determined by the last stage of the process, described by the T-model potential with $\alpha = 2/3$ \rf{Tmod}. However, the cascade regime is very interesting from the point of view of the theory of initial conditions for inflation. 

Indeed, suppose that the parameter $M$ describing the disk interactions takes the simplest value $M= O(1)$ in Planck units. Then the height of the upper potential will be Planckian, which allows to solve the problem of initial conditions for inflation in the simplest possible way, as described in~\cite{Linde:2005ht}. The Planck-size universe can be born with the scalar fields $\phi_1$ and $\phi_2$ at an infinite plateau with $V = m^{2}+M^{2} =O(1)$. According to~\cite{Linde:2005ht}, the probability of this process is not expected to be exponentially suppressed. Once this happens, the cascade inflation begins, with observational predictions determined by the last stage of the process, matching the latest observational data. 


A more general solution to the problem of initial conditions for inflation, which applies to all models discussed in our paper, can be found in  \cite{Carrasco:2015rva,East:2015ggf}.
We hope to return to a more detailed discussion of the cascade inflation in a separate publication.

\subsection{Seven-disk merger model}

Finally, we briefly discuss the possible merger of several disks. Consider for instance, 
\begin{align}
{\cal G}=&\log W_0^2-\frac12\sum_{i=1}^7\log\frac{(1-Z_i\overline{Z}_i)^2}{(1-Z_i^2)(1-\overline{Z}_i^2)}+S+\overline{S}+\cG_{S\overline S}S\overline{S},\\
\cG^{S\overline S}=&\frac{1}{W_0^2}(3 W_0^2 +\bf V ).
\end{align}
corresponding to seven disks with $\alpha_i = 1/3$. 
The scalar potential is \begin{align}
{\bf V}= \Lambda+ \frac{m^2}{7}\sum_i |Z_i|^2+\frac{M^2}{7^2}\sum_{1\leq i\leq j \leq 7}\Big ((Z_i+\overline{Z}_i)-(Z_j+\overline{Z}_j)\Big)^2,
\end{align}
and the last term gives the dynamical constraint $\phi_i=\phi_j$ where we have defined canonical fields as $Z_i=\tanh \frac{\phi_i+{\rm i}\theta_i}{\sqrt{2}}$. During inflation at $\phi_i=\phi_j = {\vp\over \sqrt 7}$, the scalar potential reads
\be
{\bf V}(\vp)= \Lambda+ {m^2} \tanh^2 {\varphi \over \sqrt {14}},
\ee
in terms of the canonically normalized inflaton field. 

The axionic directions are stabilized at their origin, and their masses are given by
\begin{align}
m_{\theta_i}^2=2(m^2+2W_0^2)-\frac{1}{7}m^2\left(7+6\cosh\sqrt{\frac27}\varphi\right)\cosh^{-4}\frac{\varphi}{\sqrt{14}}.
\end{align}
The first two constant part dominate the mass and the remaining negative part is suppressed during inflation. At the minimum, the mass of the axions becomes $m_{\theta_i}^2=\frac17 m^2+4W_0^2$ and is still positive. 

For real directions $\{\phi_i\}$, the following canonical mass eigenbasis is useful, $\varphi=\frac{1}{\sqrt7}\sum_{i=1}^7\phi_i$, and $\chi_i=\frac{1}{\sqrt{8-i}}((7-i)\phi_i-\phi_{i+1}\cdots-\phi_7)$. The inflaton is $\varphi$ and moduli $\chi_i$ are stabilized at their origin with the mass
\begin{align}
m_{\chi_i}^2=\frac{1}{7}\left(2m^2+4M^2-m^2\cosh\sqrt{\frac27}\varphi\right)\cosh^{-4}\frac{\varphi}{\sqrt{14}}.
\end{align}
As the two disk models, the mass of the moduli $\chi_i$ becomes small, and when $4M^2<m^2\cosh\sqrt{\frac27}\varphi$, they becomes tachyonic. At the minimum $\varphi=0$, the inflaton and moduli mass are given by
\begin{align}
m_\phi^2=\frac{1}{7}m^2,\qquad m_{\chi_i}^2=\frac{1}{7}m^2+\frac47M^2.
\end{align}
Note that SUSY breaking takes place at the minimum; $\cG_S=1$ and $\sqrt{\cG_S\cG^{S\overline S}\cG_{\overline S}}=\sqrt{3}W_0$.
Here again we see the advantage of using the new geometric class of models comparative to  the earlier version of the seven-disk model in Ref.~\cite{Kallosh:2017ced} where we only studied an inflationary stage.

In the seven-disk models we expect a cascade inflation with a  rich structure  due to the  multiplicity of different inflationary plateaus.  The different possibilities arise from the possible sign choices for the seven moduli. For instance, one can either take four positive and three negative, in which case 12 out of the 21 mass terms contribute. Similarly, one can have five and two, with ten mass terms etc. From this logic it follows that the  potential at the dS plateaus may take 4 different values: $V = \Lambda +m^2 +  {16 n M^2\over 7^2} $, where $n$ can be 0, 6, 10, or 12. 

\section{Discussion}

It has been realized during the last few years that both the construction of de Sitter vacua in string theory as well as building inflationary models is facilitated by the concept of an upliting $\overline {D3}$ brane. The positive energy contribution sourced by a $\overline {D3}$ brane in effective supergravity models is represented by a nilpotent multiplet $S$. Supersymmetry is spontaneously broken during inflation as well as at the exit from inflation, at the minimum of the potential, and never restored in the class of models we described here:  $\overline {D3}$ brane induced geometric inflationary models. 

The effective supergravity of these models is  described by the geometry of the CY moduli, $\cG_0(T^i, \overline T^i)$ and by the geometry of the nilpotent superfield $\cG_{S \overline S} (T^i, \overline T^i)S\overline S$.  In our models it
is given by the expression
\be
\cG (T^i, \overline T^i; S, \overline S)= \cG_0(T^i, \overline T^i) + S +\overline S + \cG_{S\overline S} (T^i, \overline T^i)S\overline S.
\label{calG1}\ee
Subject to specific assumptions about the geometry of $T^i$ moduli \eqref{K}, satisfied by simple examples like a shift symmetric canonical geometry \eqref{shift} or a disk geometry \eqref{disk}, one finds a simple relation between the inflationary potential  and geometry of the $\overline {D3}$ brane in the background of the $T^i$ fields:
 \begin{align}
\cG^{S \overline S} (T^i, \overline T^i)  =   { \mathbf{ V} (T^i, \overline T^i)+3 |m_{3/2}|^2 \over   |m_{3/2}|^2}     .
 \label{new1}  \end{align} 
 This relation leads to a model building procedure of the following kind. Once the desired potential $ \mathbf{ V} (T^i, \overline{ T}^i)$ is determined, one can use the relation 
 \rf{new1} to produce the geometry $\cG^{S \overline S} (T^i,\overline{ T}^i)$. The remaining problem for each choice of inflationary model is to check that all non-inflaton directions are stabilized.
 
We have found  that such a procedure leads to  rather simple models with desirable properties. In particular, in models with  one modulus $T$ one finds that axions are stable during inflation. At the minimum, the masses of the inflaton and axion also tend to be positive for the appropriate choices of the potentials where there is an exit from inflation at the minimum of the potential. These desirable stability properties of the potential are in a nice agreement with the positivity of the $S$-field metric $\cG_{S \overline S} (T^i, \overline{ T}^i)$.

Our examples illustrate the main result of the paper: we build desirable cosmological models with inflationary potentials $  \mathbf{ V} (T^i,\overline{ T}^i)$ which are in agreement with the data, and we `read from the sky' the geometry of the $\overline {D3}$ brane in CY bulk supporting these models as shown in eq.~\rf{new1}. The geometric nature of all these models manifests itself in the fact that the bisectional curvature is always present and is defined by the slow-roll parameters as shown in Sec.~\ref{curvature}. At the exit from inflation at the minimum this curvature gives  a positive contribution to the masses.
 
We find that this geometric formulation of effective supergravity inflationary models inspired by string theory is the most powerful tool  for model building. Their first advantage is that they are easily associated with string theory due to fundamental role of the uplifting $\overline {D3}$ brane, interacting with other moduli. The second advantage is that for specific choices of \K\, geometries of the moduli fields $T^i$, the only input comes from the nilpotent field geometry, $\cG_{S\overline S}(T^i, \overline{ T}^i)$, related to the potential. In previously existing models with generic superpotential $W= Sf(T^i) + g(T^i)$, the main input is via two holomorphic functions $f(T^i)$ and $ g(T^i)$, which should satisfy additional constraints. This made the model building more involved than in the approach developed in this paper.
   The third advantage is the fact that, by construction, the nilpotency condition $F_S\neq 0$ is  satisfied everywhere, including the minimum of the potential.  The mere existence of the uplifting $\overline {D3}$ brane interacting with the bulk geometry means that supersymmetry is nonlinearly realized and always spontaneously
  broken.
  
In conclusion, the new cosmological models, $\overline {D3}$ induced geometric models, defined by a geometric \K\, function in eq.~\rf{calG1}, lead to simple dynamical cosmological models of the inflationary evolution of the space-time, based on the geometry of the scalar manifold. The dynamics of these models is the consequence of their geometry.

\
  
 \noindent{\bf {Acknowledgments:}} We are grateful to E. Bergshoeff, K. Dasgupta, S. Ferrara, D. Freedman, S. Kachru, E. McDonough, M. Scalisi, F. Quevedo, A. Uranga,  A. Van Proeyen, A. Westphal, and T. Wrase   for stimulating discussions and collaborations on related work.  The work  of  RK, AL and YY is supported by SITP and by the US National Science Foundation grant PHY-1316699.  The work of AL is also supported by the Templeton foundation grant ``Inflation, the Multiverse, and Holography''. DR is grateful to SITP for the hospitality when this work was initiated.
 \

\end{document}